\documentclass[12pt]{iopart}

\usepackage{graphicx}
\usepackage{epsfig}    

\usepackage[footnotesize,bf]{caption}	%Abb...fett gedruckt 

\begin{document}
%%%%%%%%%%%%%%%%%%%%%%%%%%%%%%%%%%%%%%%%%%%%%%%%%%%%%%%%%%%%%%%%%%%%%%%%%%%%%%
\title{Lambda production in central Pb+Pb collisions at CERN-SPS energies} 

\author{A. Mischke for the NA49 Collaboration}

\address{Gesellschaft f\"ur Schwerionenforschung, D-64291 Darmstadt, Germany}  

\ead{a.mischke@gsi.de}

%%%%%%%%%%%%%%%%%%%%%%%%%%%%%%%%%%%%%%%%%%%%%%%%%%%%%%%%%%%%%%%%%%%%%%%%%%%%%%
%                            author list
%%%%%%%%%%%%%%%%%%%%%%%%%%%%%%%%%%%%%%%%%%%%%%%%%%%%%%%%%%%%%%%%%%%%%%%%%%%%%%
\vspace{0.75cm}

\noindent
S.V.~Afanasiev$^{9}$,T.~Anticic$^{20}$, D.~Barna$^{5}$,
J.~Bartke$^{7}$, R.A.~Barton$^{3}$,
L.~Betev$^{10}$, \mbox{H.~Bia{\l}\-kowska$^{17}$}, A.~Billmeier$^{10}$,
C.~Blume$^{8}$, C.O.~Blyth$^{3}$, B.~Boimska$^{17}$, M.~Botje$^{1}$,
J.~Bracinik$^{4}$, R.~Bramm$^{10}$, R.~Brun$^{11}$,
P.~Bun\v{c}i\'{c}$^{10,11}$, V.~Cerny$^{4}$,
J.G.~Cramer$^{19}$, P.~Csat\'{o}$^{5}$, P.~Dinkelaker$^{10}$,
V.~Eckardt$^{16}$, P.~Filip$^{16}$,
H.G.~Fischer$^{11}$, Z.~Fodor$^{5}$, P.~Foka$^{8}$, P.~Freund$^{16}$,
V.~Friese$^{15}$, J.~G\'{a}l$^{5}$,
M.~Ga\'zdzicki$^{10}$, G.~Georgopoulos$^{2}$, E.~G{\l}adysz$^{7}$, 
S.~Hegyi$^{5}$, C.~H\"{o}hne$^{15}$, G.~Igo$^{14}$,
P.G.~Jones$^{3}$, K.~Kadija$^{11,20}$, A.~Karev$^{16}$,
V.I.~Kolesnikov$^{9}$, T.~Kollegger$^{10}$, M.~Kowalski$^{7}$, 
I.~Kraus$^{8}$, M.~Kreps$^{4}$, M.~van~Leeuwen$^{1}$, 
P.~L\'{e}vai$^{5}$, A.I.~Malakhov$^{9}$, S.~Margetis$^{13}$,
C.~Markert$^{8}$, B.W.~Mayes$^{12}$, G.L.~Melkumov$^{9}$,
J.~Moln\'{a}r$^{5}$, J.M.~Nelson$^{3}$,
G.~P\'{a}lla$^{5}$, A.D.~Panagiotou$^{2}$,
K.~Perl$^{18}$, A.~Petridis$^{2}$, M.~Pikna$^{4}$, L.~Pinsky$^{12}$,
F.~P\"{u}hlhofer$^{15}$,
J.G.~Reid$^{19}$, R.~Renfordt$^{10}$, W.~Retyk$^{18}$,
C.~Roland$^{6}$, G.~Roland$^{6}$, A.~Rybicki$^{7}$, T.~Sammer$^{16}$,
A.~Sandoval$^{8}$, H.~Sann$^{8}$, N.~Schmitz$^{16}$, P.~Seyboth$^{16}$,
F.~Sikl\'{e}r$^{5}$, B.~Sitar$^{4}$, E.~Skrzypczak$^{18}$,
G.T.A.~Squier$^{3}$, R.~Stock$^{10}$, H.~Str\"{o}bele$^{10}$, T.~Susa$^{20}$,
I.~Szentp\'{e}tery$^{5}$, J.~Sziklai$^{5}$,
T.A.~Trainor$^{19}$, D.~Varga$^{5}$, M.~Vassiliou$^{2}$,
G.I.~Veres$^{5}$, G.~Vesztergombi$^{5}$,
D.~Vrani\'{c}$^{8}$, S.~Wenig$^{11}$, A.~Wetzler$^{10}$, C.~Whitten$^{14}$,
I.K.~Yoo$^{15}$, J.~Zaranek$^{10}$, J.~Zim\'{a}nyi$^{5}$

\vspace{0.5cm}
\begin{footnotesize}
\noindent
$^{1}$NIKHEF, Amsterdam, Netherlands. \\
$^{2}$Department of Physics, University of Athens, Athens, Greece.\\
$^{3}$Birmingham University, Birmingham, England.\\
$^{4}$Comenius University, Bratislava, Slovakia.\\
$^{5}$KFKI Research Institute for Particle and Nuclear Physics, Budapest, Hungary.\\
$^{6}$MIT, Cambridge, USA.\\
$^{7}$Institute of Nuclear Physics, Cracow, Poland.\\
$^{8}$Gesellschaft f\"{u}r Schwerionenforschung (GSI), Darmstadt, Germany.\\
$^{9}$Joint Institute for Nuclear Research, Dubna, Russia.\\
$^{10}$Fachbereich Physik der Universit\"{a}t, Frankfurt, Germany.\\
$^{11}$CERN, Geneva, Switzerland.\\
$^{12}$University of Houston, Houston, TX, USA.\\
$^{13}$Kent State University, Kent, OH, USA.\\
$^{14}$University of California at Los Angeles, Los Angeles, USA.\\
$^{15}$Fachbereich Physik der Universit\"{a}t, Marburg, Germany.\\
$^{16}$Max-Planck-Institut f\"{u}r Physik, Munich, Germany.\\
$^{17}$Institute for Nuclear Studies, Warsaw, Poland.\\
$^{18}$Institute for Experimental Physics, University of Warsaw, Warsaw, Poland.\\
$^{19}$Nuclear Physics Laboratory, University of Washington, Seattle, WA, USA.\\
$^{20}$Rudjer Boskovic Institute, Zagreb, Croatia.\\
\end{footnotesize}

%%%%%%%%%%%%%%%%%%%%%%%%%%%%%%%%%%%%%%%%%%%%%%%%%%%%%%%%%%%%%%%%%%%%%%%%%%%%%%
%                                abstract
%%%%%%%%%%%%%%%%%%%%%%%%%%%%%%%%%%%%%%%%%%%%%%%%%%%%%%%%%%%%%%%%%%%%%%%%%%%%%%
\begin{abstract}

\noindent
In this paper we present recent results from the NA49 experiment for 
$\Lambda$ and $\bar{\Lambda}$ hyperons produced in central Pb+Pb
collisions at 40, 80 and 158~A$\cdot$GeV. 
Transverse mass spectra and rapidity distributions for $\Lambda$
are shown for all three energies. The shape of the rapidity
distribution becomes flatter with increasing beam energy. The
multiplicities at mid-rapidity as well as the total yields are
studied as a function of collision energy including AGS
measurements. The ratio $\Lambda/\pi$ at mid-rapidity and in 4$\pi$
has a maximum around 40 A$\cdot$GeV.  
In addition, $\bar{\Lambda}$ rapidity distributions have been measured
at 40 and 80~A$\cdot$GeV, which allows to study the
$\bar{\Lambda}$/$\Lambda$ ratio.  

\end{abstract}

\vspace{-0.4cm}

\submitto{\JPG}
\pacs{appear here}

%\maketitle

%\setlength{\baselineskip}{1.2cm}       %%%%%ZEILENABSTAND%%%%

%\vspace{-12cm}
%
%%%%%%%%%%%%%%%%%%%%%%%%%%%%%%%%%%%%%%%%%%%%%%%%%%%%%%%%%%%%%%%%%%%%%%%%%%%%%%
%
\section{Introduction}
Relativistic nucleus-nucleus collisions allow the investigation of
nuclear matter at high temperatures and densities. 
In-medium production of strangeness in nucleus-nucleus collisions is
expected to differ from what is known from elementary hadron-hadron
interactions. 
The measurement of strange baryons like $\Lambda$(uds) hyperons, which
contain between 25 and 60$\%$ of the total strangeness produced
(depending on the energy), offer the possibility to study
simultaneously strangeness production and the effect of baryon density
in A-A collisions.
The excitation function of hyperon production will probe the behavior
of strange baryons at various energy densities in the interaction zone,
which depend on the collision energy.  
The NA49 collaboration contributes to the excitation function the
measurements of $\Lambda$ and $\bar{\Lambda}$ hyperons in central
Pb+Pb collisions at 40, 80 and 158 A$\cdot$GeV. 

%At energy densities of a few GeV per $fm^3$ the hadrons cannot exists
%as bounded states, they probably resolve into a plasma of quarks and
%gluons. Lattice QCD considerations predict such a Quark-Gluon-Plasma
%state, whereas strangeness.

%... The $\Lambda$ measurement is specific interest because of the
%predicted energy dependence of the strangeness over entropy ratio
%which can also be studied in the $\Lambda/\pi$

%
%%%%%%%%%%%%%%%%%%%%%%%%%%%%%%%%%%%%%%%%%%%%%%%%%%%%%%%%%%%%%%%%%%%%%%%%%%%%%%
%
\section{The NA49 experiment at CERN-SPS}
\begin{figure}[h]
 \begin{minipage}[b]{15cm}
  \begin{center}
   \includegraphics[height=5cm]{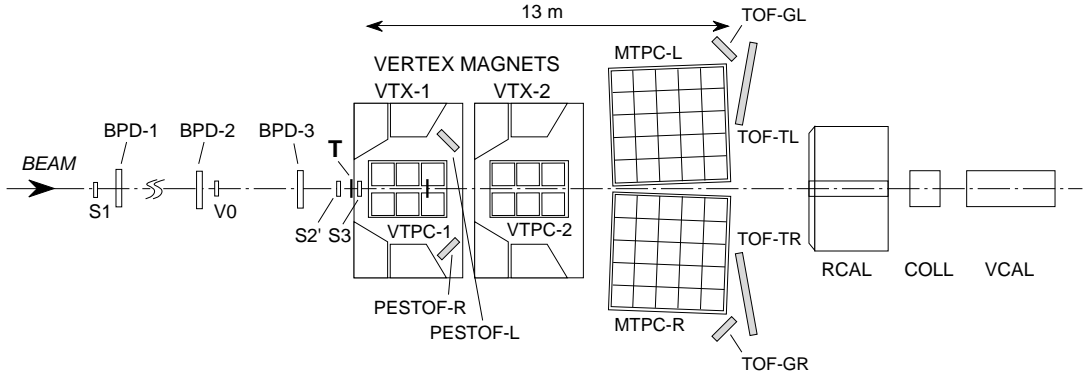}
  \end{center}
  \vspace{-0.5cm}
  \caption{\protect 
Schematic arrangement of the experiment NA49~\cite{NA49_Det}. For
details see text.}	
  \label{abb1}
 \end{minipage}  
\end{figure}

The experiment NA49 is a large acceptance hadron spectrometer at the
CERN-SPS (figure~\ref{abb1}). Tracking and particle identification by
the measurement of the specific energy loss (d$E$/d$x$)
is performed by two Time Projection Chambers (VTPC-1 and VTPC-2) located
inside two vertex magnets (1.5 and 1.1~T, respectively) and two large
volume TPC's (MTPC-L and MTPC-R) situated downstream of the magnets
symmetrically to the beam line~\cite{NA49_Det}. The relative d$E$/d$x$
resolution is 3-4~$\%$ and the momentum resolution
$\frac{\sigma(p)}{p^2}= 0.3\cdot10^{-4}(\rm{GeV}/c)^{-1}$. 
Two {\mbox Time-of-Flight} walls (TOF-TL and TOF-TR) give additional
particle identification near {\mbox mid-rapidity}. 
The trigger on centrality is based on the energy measurement in the
forward calorimeter (VCAL), which detects the projectile spectator
nucleons.  

%
% The exp. set-up, shown schematically in fig1, is described in detail
% in ref1...
%
%%%%%%%%%%%%%%%%%%%%%%%%%%%%%%%%%%%%%%%%%%%%%%%%%%%%%%%%%%%%%%%%%%%%%%%%%%%%%%
%
\section{$\Lambda$ production}
The NA49 collaboration has taken central Pb+Pb data at 40, 80 and
158~GeV per {\mbox nucleon}. For the present analysis, the 7~$\%$
most central interactions at 40 and 80~A$\cdot$GeV were selected. 
The resulting event sample has 349 participating nucleons on average. 
For 158~A$\cdot$GeV the trigger selected the 10~$\%$ most 
central events (335 participants).
The number of analyzed events is 400k for 40 and 158~A$\cdot$GeV as well
as 300k for 80~A$\cdot$GeV.

\begin{figure}[h]
\begin{minipage}[b]{5cm}
 \begin{center}
 \includegraphics[height=8.cm]{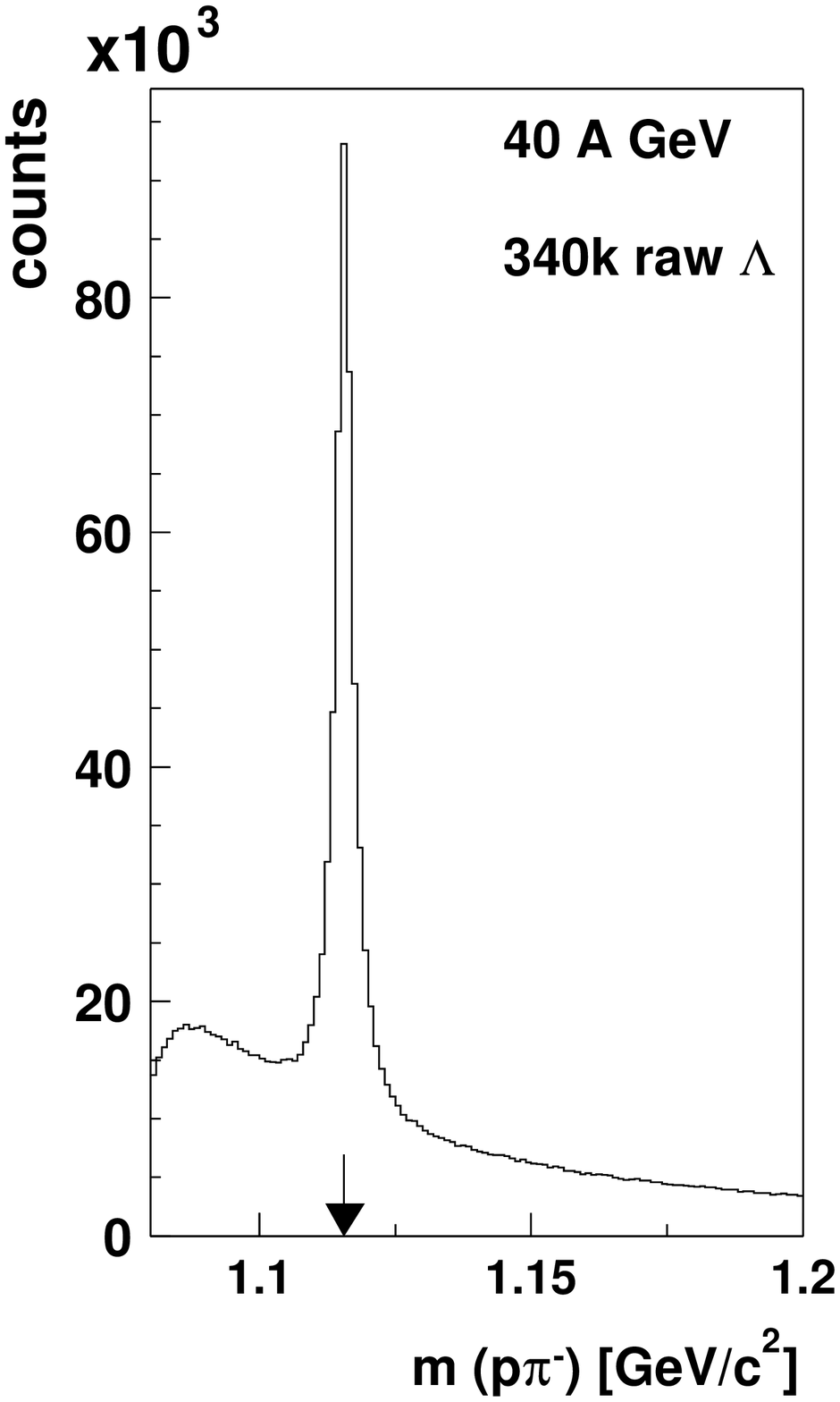}
 \end{center}
\end{minipage} 
\hspace{\fill}
\begin{minipage}[b]{5cm}
 \begin{center}
 \includegraphics[height=8.cm]{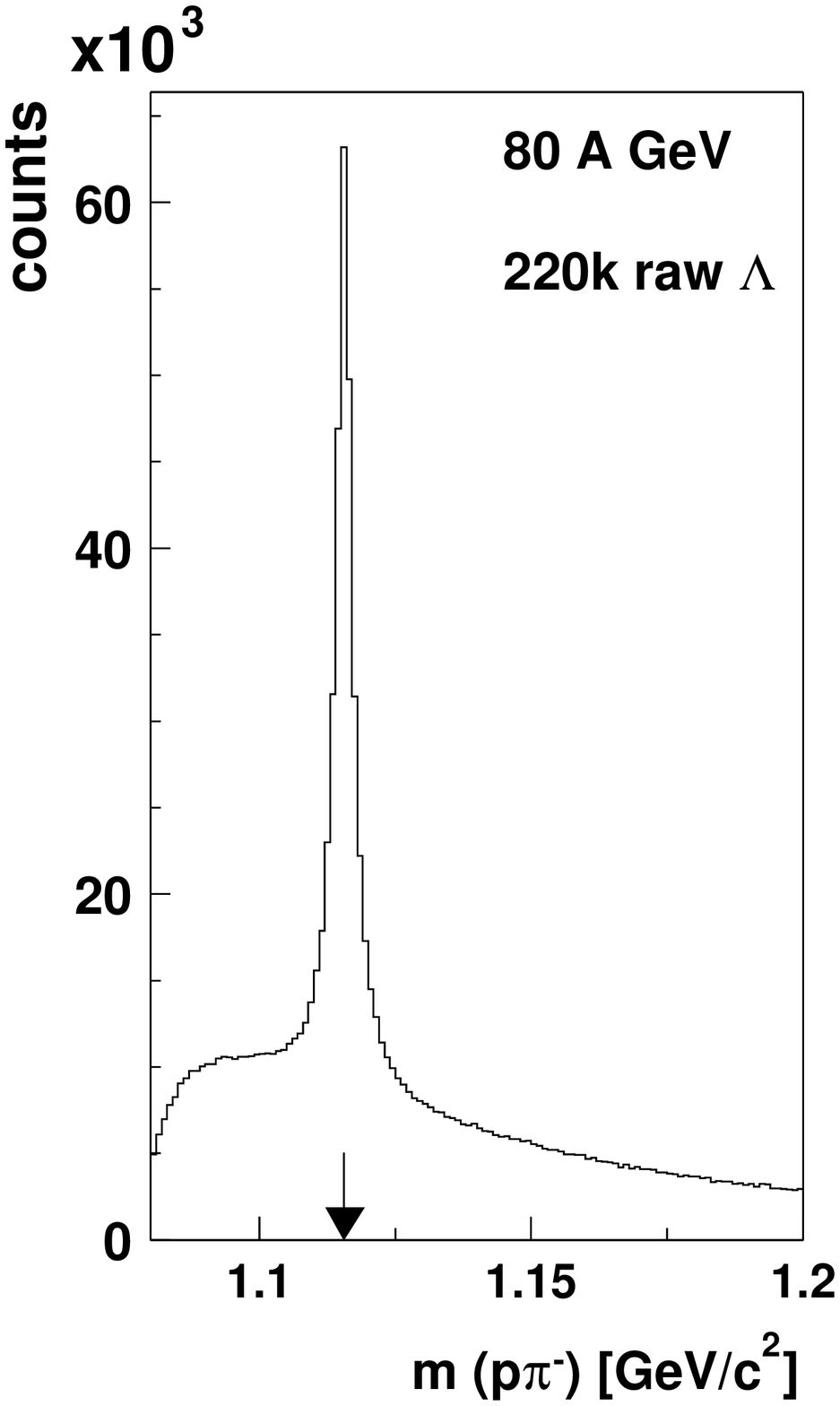}
 \end{center}
\end{minipage} 
\hspace{\fill}
\begin{minipage}[b]{5cm}
 \begin{center}
 \includegraphics[height=8.cm]{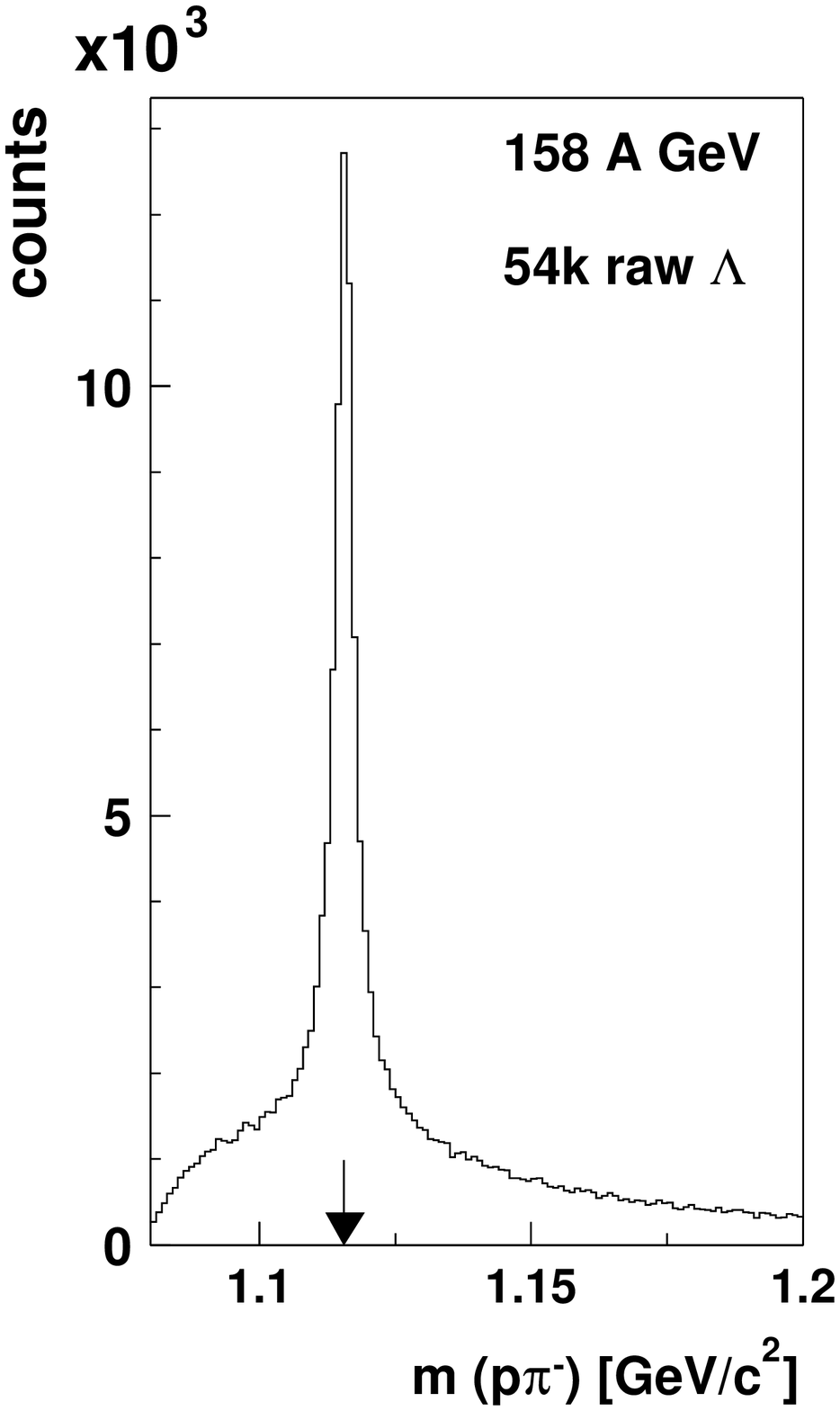}
 \end{center}
\end{minipage}  
 \vspace{-1.cm}
 \caption{\protect 
The invariant mass distributions of p-$\pi^-$ pairs in central Pb+Pb
reactions at 40~(left), 80~(middle) and 158~GeV per
nucleon~(right). The measured masses are in good agreement with the
PDG value of 1.115~GeV/$c^{2}$, indicated by the arrows. The mass
resolution is 2~MeV/$c^{2}$ for all three energies.} 
 \label{abb2a-c}
\end{figure}

$\Lambda$ and $\bar{\Lambda}$ hyperons are identified by
reconstructing their decay topologies \mbox{ $\Lambda \rightarrow
p+\pi^{-}$} and $\bar{\Lambda} \rightarrow \bar{p}+\pi^{+}$, 
respectively, with a branching fraction of 63.9$\%$. 
The charged decay products are measured with the TPC's.
% using global tracking. %%%%%%%%%% 
% A set of quality cuts was applied to reduce the background.%%%%%%%%%% 
In figure~\ref{abb2a-c}, the invariant mass distributions are shown
for all three energies. Clear signals are observed with reasonable
signal-to-background ratios. The agreement between the measured masses
and the nominal value, indicated by the arrows, is excellent. The mass
resolution is 2~MeV/$c^{2}$ ($\sigma_{\rm m}$).
The signals were extracted after subtracting the background, which is
well described by a third-order polynomial. The average numbers of
$\Lambda$ reconstructed are also shown in the figures. 

\subsection{Spectra}
\begin{figure}[h]
\begin{minipage}[b]{6.5cm}
 \begin{center}
 \includegraphics[height=6.47cm]{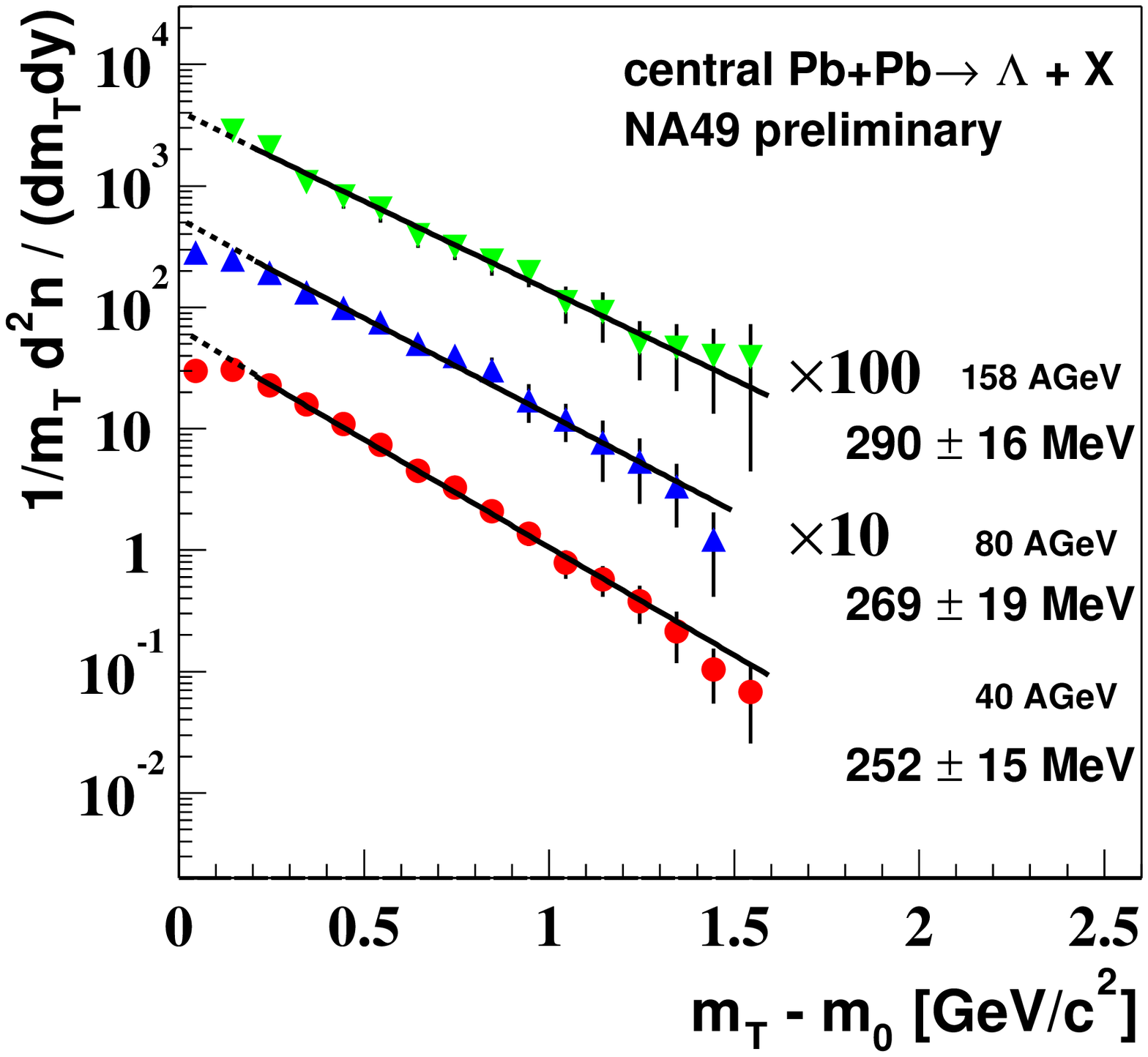}
 \end{center}
 \vspace{-1.2cm}
 \caption{\protect 
A summary of the transverse mass spectra for all three energies. The
fitted inverse slope parameter increases with increasing energy.}
 \label{abb3}
\end{minipage} 
%
%\hspace{\fill}
\hspace{0.5cm}
\begin{minipage}[b]{8cm}
 \begin{center}
 \includegraphics[height=6.3cm]{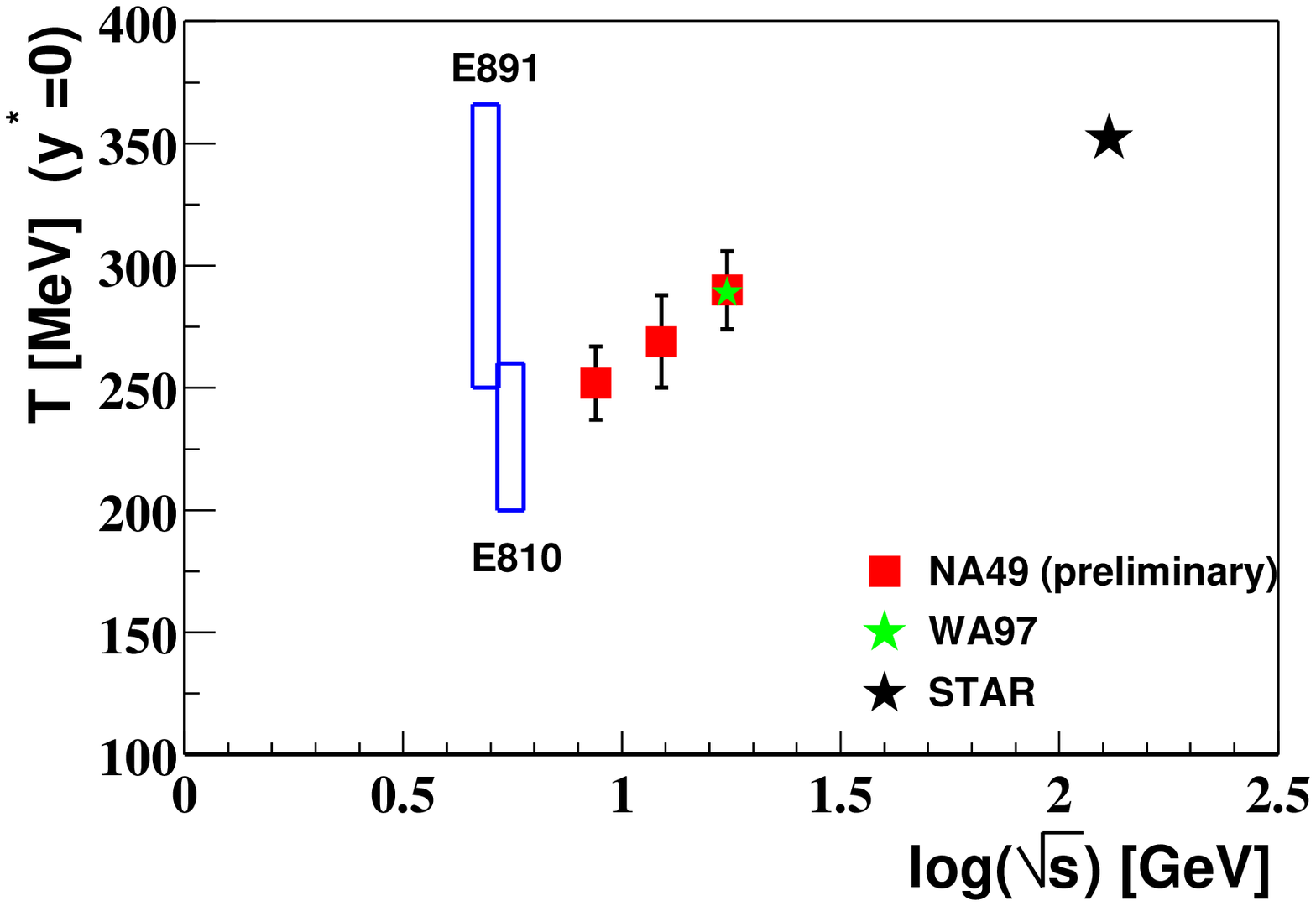}
 \end{center}
 \vspace{-1.2cm}
 \caption{\protect
The inverse slope parameter $T$ as a function of the beam energy. The
open rectangular symbols are estimates of $T$. A slight increase with 
energy is observed. Note the suppressed zero.}
 \label{abb4}
\end{minipage} 
\end{figure}

We have performed corrections bin by bin in rapidity and transverse
momentum for geometrical acceptance and tracking efficiency
using a full Monte Carlo simulation of the detector in GEANT.
The reconstruction efficiency was obtained by embedding GEANT
simulated $\Lambda$ in raw data events (one per event) followed by the
same reconstruction procedure, which was used for the normal data
sample.
The transverse mass 
($m_{\rm T}=\sqrt{p_{\rm T}^2+m_0^2}$, where $m_0$ is the rest mass of
the particle)  
distributions of $\Lambda$ (at mid-rapidity) are plotted in
figure~\ref{abb3}. All spectra follow a single exponential function in
$m_{\rm T}$ according to    

\[ \frac{1}{m_{\rm T}} \frac{d^2n}{dm_{\rm T} dy} \propto e^{- \frac{m_{\rm T}}{T}},\]

where $T$ is the inverse slope parameter. 
The slope factors are fitted in the interval 
0.2~GeV/$c^{2}$ $\le m_{\rm T} \le$ 1.6~GeV/$c^{2}$ resulting in 
$T=(290~\pm~16)$~MeV at 158~A$\cdot$GeV, 
$T=(269~\pm~19)$~MeV at 80~A$\cdot$GeV and
$T=(252~\pm~15)$~MeV at 40~A$\cdot$GeV.  
The slope parameter increases with increasing energy. This is
illustrated in figure~\ref{abb4}, where the inverse slope $T$ is shown
as a function of center-of-mass energy per nucleon nucleon pair,
$\sqrt{s}$.  
The value published by the WA97 collaboration at
158~A$\cdot$GeV~\cite{WA97_1} and the preliminary result from the STAR
collaboration (central Au+Au at $\sqrt{s}$=~130~GeV~\cite{STAR}) as
well as an estimate of the slope at AGS energies~\cite{lam_AGS_Temp}
confirm the trend seen in our data. Another estimate at AGS beam
energies~\cite{lam_topAGS1} seems to be too high but still compatible due
to its large errors.

\begin{figure}[h]
\vspace{-.5cm}
\hspace{3cm}
\begin{minipage}[b]{9.5cm}
 \begin{center}
 \includegraphics[height=6.cm]{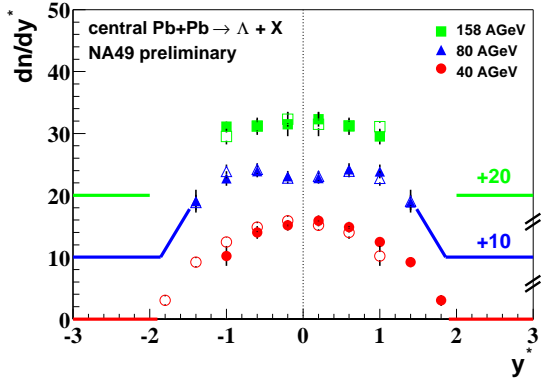}
 \end{center}
 \vspace{-1.2cm}
 \caption{\protect 
The rapidity distributions of $\Lambda$ hyperons in central Pb+Pb
collisions at 40, 80 and 158~A$\cdot$GeV. The baseline for 80 and
158~A$\cdot$GeV is shifted by 10 and 20 units, respectively.}
 \label{abb5}
\end{minipage} 
\end{figure}

At 40 and 80~A$\cdot$GeV the acceptance covers the full transverse
momentum region down to $p_{\rm T}$~=~0~GeV/$c$. The yield in each
rapidity bin is the integral of the $p_{\rm T}$-distribution. At
158~A$\cdot$GeV, however, an extrapolation using the fitted slope
parameter is necessary to determine the yield (extrapolation factor =
1.22).
Feeddown corrections for cascade decays are not applied, but these
corrections are expected to be small.
The resulting rapidity distributions for all three energies are
compared in figure~\ref{abb5}. The baseline for 40~A$\cdot$GeV is at
0, indicated by the horizontal line.  
For clearness, the baseline for 80 and 158~A$\cdot$GeV is shifted by
10 and 20, respectively. The reflected points (open circles) are close
to the measured ones (filled circles). At 40~A$\cdot$GeV the produced 
$\Lambda$ are concentrated at mid-rapidity, whereas the distribution
becomes flatter at higher energies.  

\subsection{Energy dependence}
\begin{figure}[h]
\begin{minipage}[b]{7cm}
 \begin{center}
 \includegraphics[height=6.cm]{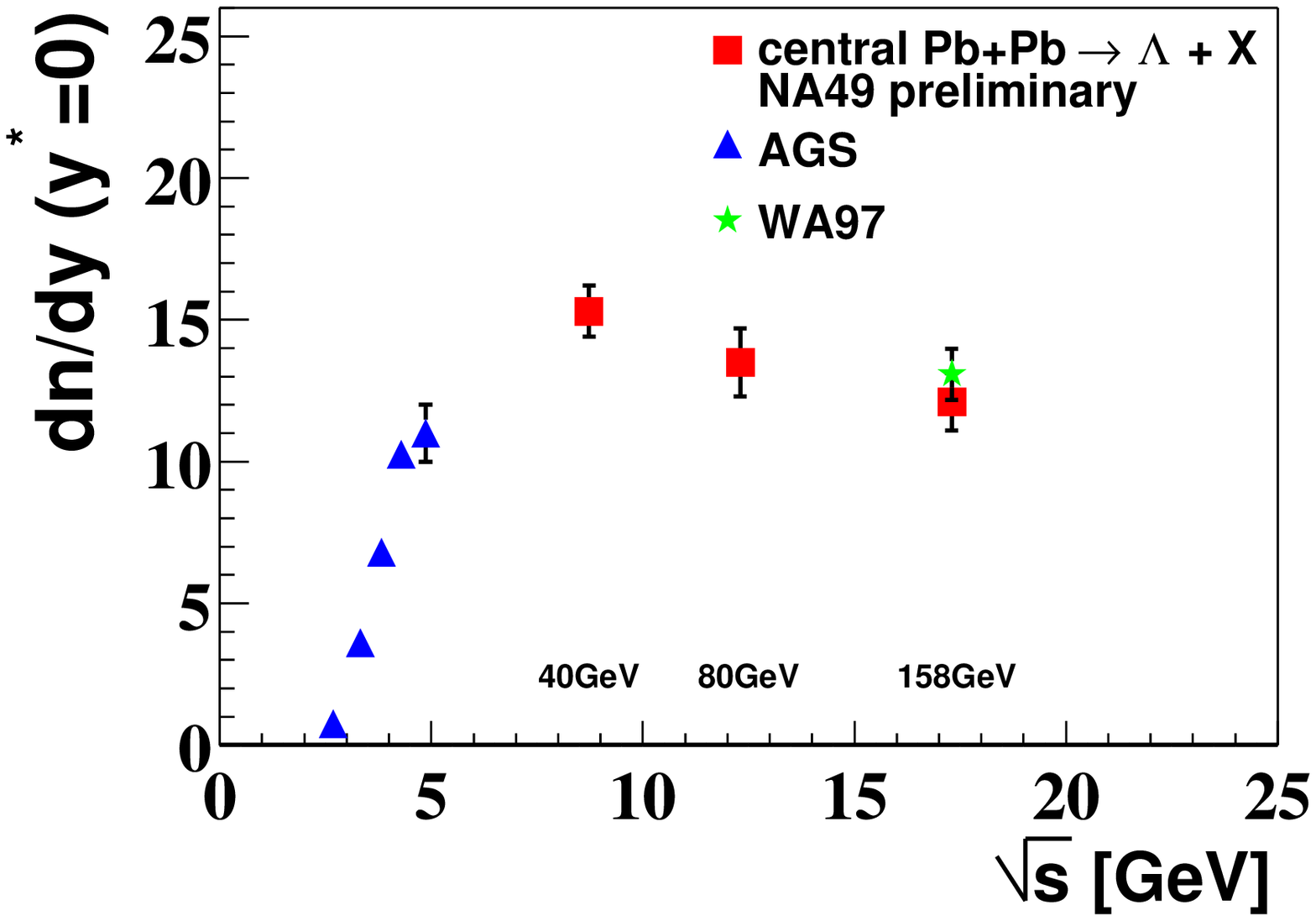}
 \end{center}
\end{minipage} 
\hspace{1.cm}
\begin{minipage}[b]{7cm}
 \begin{center}
 \includegraphics[height=6.cm]{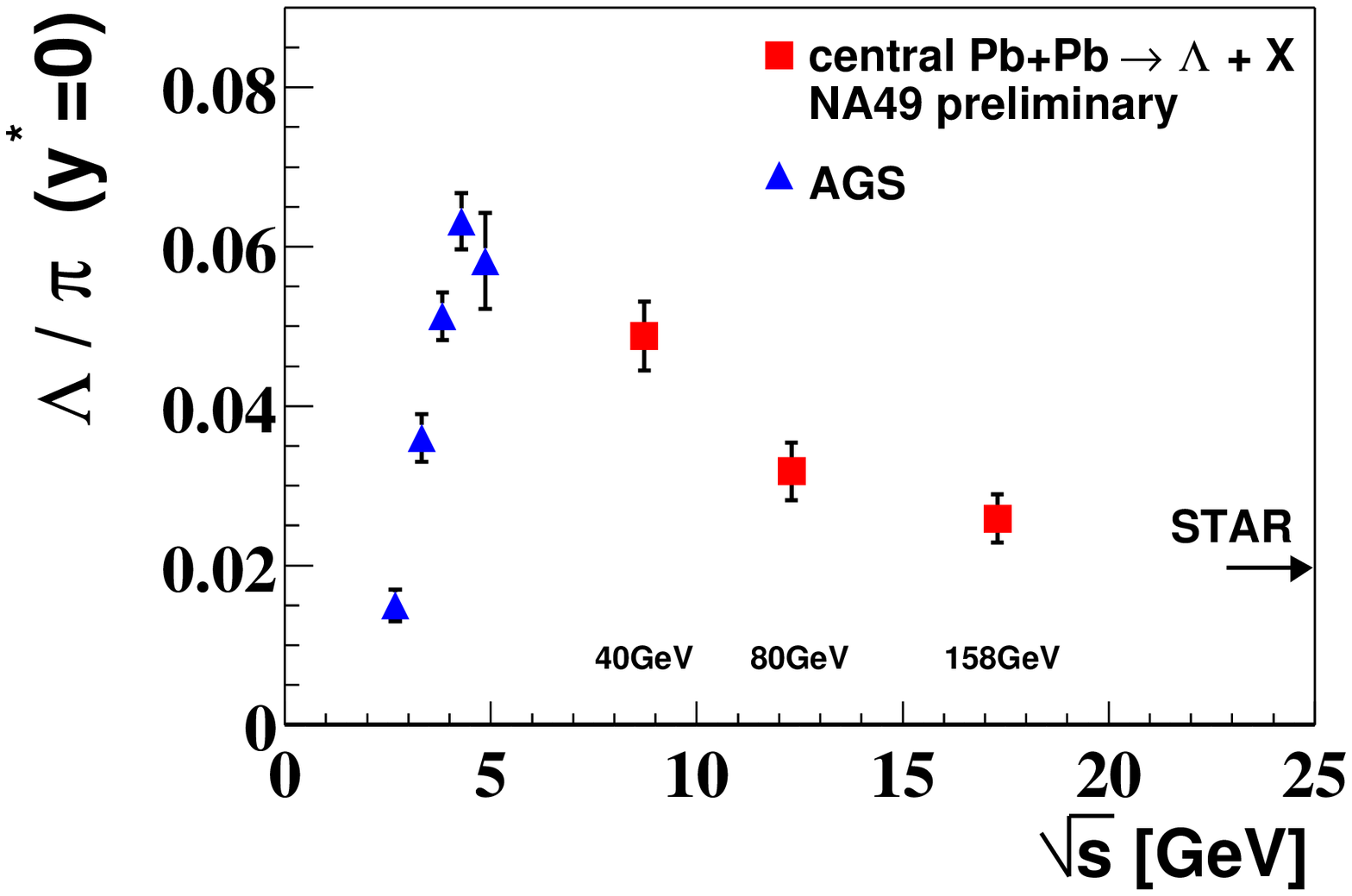}
 \end{center}
\end{minipage} 
 \vspace{-1.cm}
 \caption{\protect 
(left)~The rapidity density as a function of c.m. energy.
(right)~The energy dependence of the $\Lambda/\pi$ ratio (at
mid-rapidity). For comparison, AGS measurements are also shown. The
first result from the STAR collaboration~\cite{STAR}, measured at
$\sqrt{s}$=~130~GeV, is illustrated with a horizontal arrow.}
 \label{abb6}
\end{figure}

The maximum of the $\Lambda$ rapidity distribution decreases with
increasing collision energy, as shown in figure~\ref{abb5}. This is
highlighted in figure~\ref{abb6}~(left), where the yield at mid-rapidity
($y^{\star}$=0) is plotted as a function of c.m.~energy, including
measurements from AGS energies~\cite{lam_topAGS1}~\cite{lam_AGS1} and 
from the WA97 collaboration~\cite{WA97_2}.  
The rapidity density ${\frac{dn}{dy}}(y^{\star}=0)$ 
steeply increases
at low energies and reaches its maximum at or below 40~A$\cdot$GeV.   
In the ratio $\Lambda/\pi$ (see figure~\ref{abb6} right) this effect
is even enhanced. 
The pion multiplicity is calculated in the form $\pi = 1.5 \cdot
(\pi^+ + \pi^-)$, using measurements from~\cite{pions} and assuming
isospin symmetry of the interacting nuclei. 
The ratio $\Lambda/\pi$ shows a steep increase at AGS energies and a
strong drop at SPS energies. Much further on the energy scale, at
$\sqrt{s}$=~130~GeV, the first result from the STAR
collaboration~\cite{STAR}, indicated by the arrow, fits smoothly into
this systematics. The strong non-monotonic energy behavior is
explained by the superposition of the slightly decreasing $\Lambda$
yield and the increasing $\pi$ yield in the range of SPS energies.  

\begin{figure}[h]
\begin{minipage}[b]{7cm}
 \begin{center}
 \includegraphics[height=6.cm]{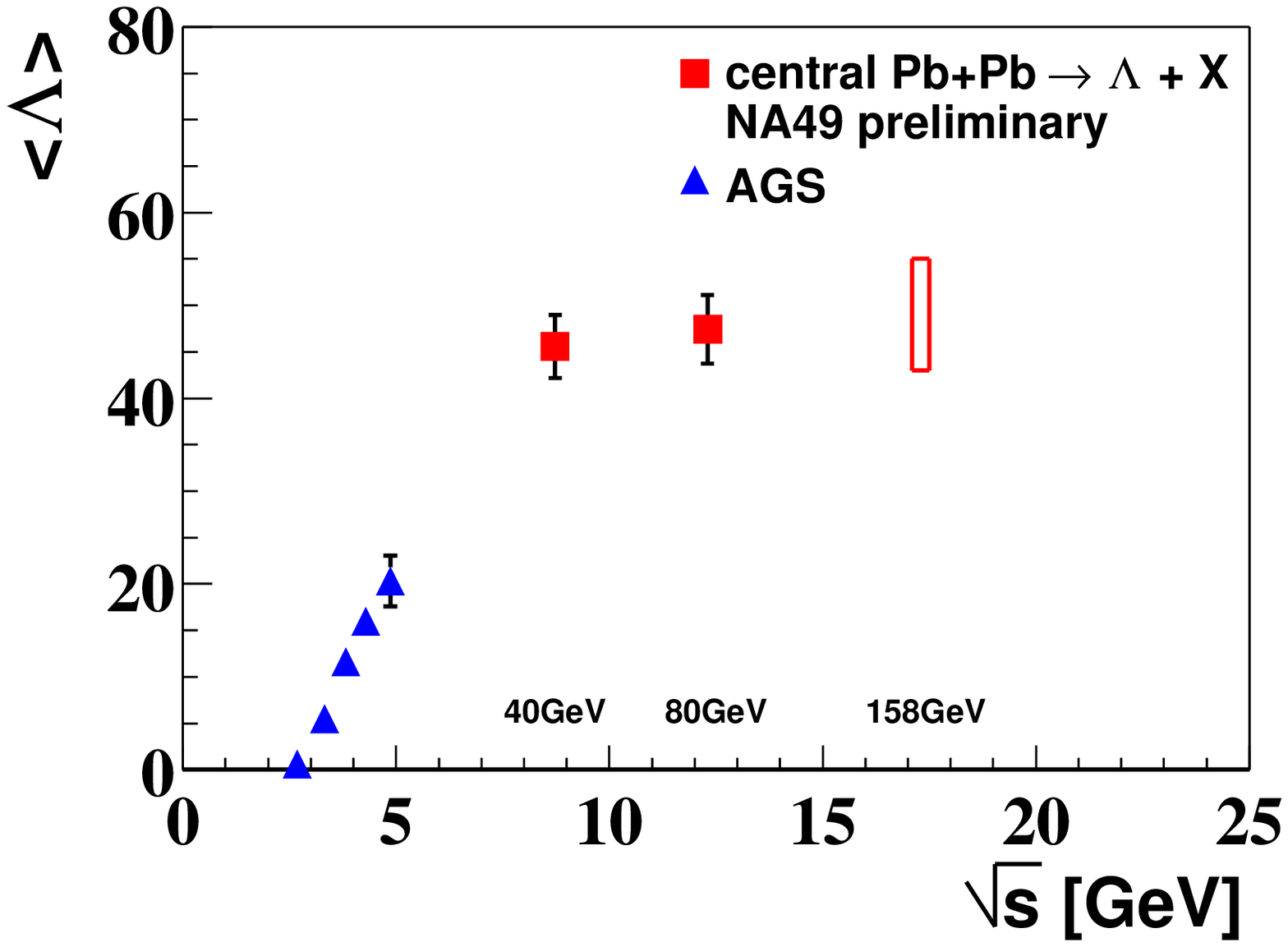}
 \end{center}
\end{minipage} 
\hspace{1cm}
\begin{minipage}[b]{7cm}
 \begin{center}
 \includegraphics[height=6.cm]{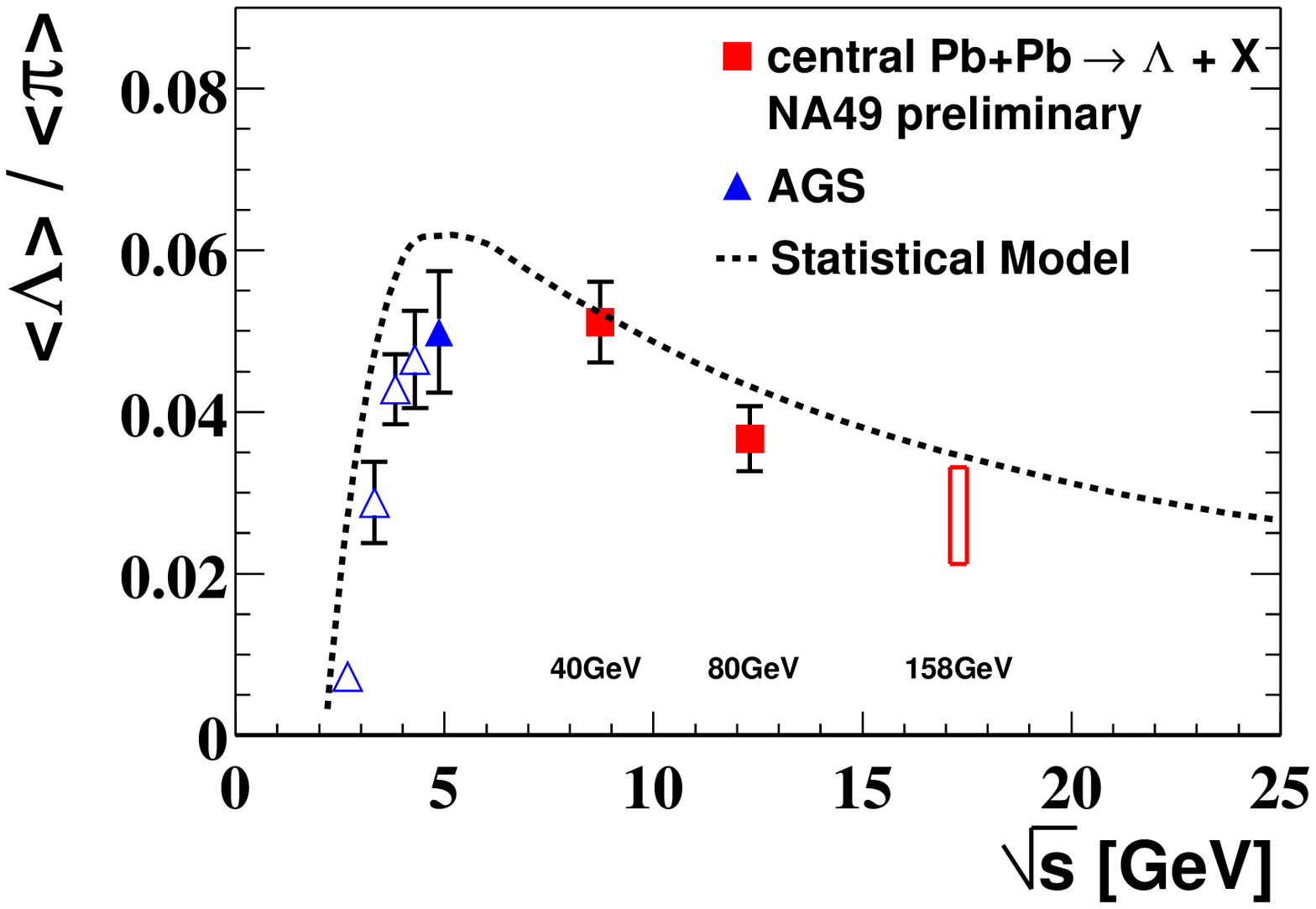}
 \end{center}
\end{minipage} 
 \vspace{-1.cm} 
 \caption{\protect 
The total $\Lambda$ multiplicity~(left) and the ratio of total
$\Lambda$ and $\pi$ yields~(right) as a function of c.m. energy for
central Pb+Pb collisions. The dotted line is a statistical
model calculation by~\cite{model2}.}     
 \label{abb7}
\end{figure}

The large acceptance of the NA49 detector allows to study also the
particle yields in the full phase space, as illustrated in
figure~\ref{abb5}.  
The total yields are obtained by integration of $\frac{dn}{dy}$ with
only small extrapolations into unmeasured regions at 40 and
80~A$\cdot$GeV.   
For 158~A$\cdot$GeV the rapidity distribution must be extrapolated,
using realistic estimates of the tails of the $\frac{dn}{dy}$
distribution, e.g. those of the net-proton distribution at
158~A$\cdot$GeV~\cite{net_proton} or the $\Lambda$ rapidity
distribution in central S+S collisions~\cite{NA35}.  
In figure~\ref{abb7}~(left), the total multiplicities as well as the AGS
measurements~\cite{lam_AGS2}~\cite{lam_topAGS2} are summarized. 
The total $\Lambda$ multiplicity increases in the AGS region and
saturates at SPS energies.  
The $\langle\Lambda\rangle / \langle\pi\rangle$ ratio (the pion yields
are taken from~\cite{pions}) follows a non-monotonic energy dependence
(see figure~\ref{abb7} right). 

The non-monotonic behavior of the strangeness-to-entropy ratio is
predicted to occur in the case of deconfinement~\cite{model1}. The
$\langle\Lambda\rangle / \langle\pi\rangle$ ratio is sensitive to the
strangeness-to-entropy ratio and in addition to the baryon
density. The effect of baryon density is strong as shown in
figure~\ref{abb7}~(right), where our results are compared with 
statistical model calculations by~\cite{model2}, which are
represented by the dotted line. The curve lies systematically above
the measured points.

%
%%%%%%%%%%%%%%%%%%%%%%%%%%%%%%%%%%%%%%%%%%%%%%%%%%%%%%%%%%%%%%%%%%%%%%%%%%%%%%
%
\section{$\bar{\Lambda}$ production at 40 and 80~A$\cdot$GeV}

\begin{figure}[h]
\begin{minipage}[b]{7cm}
 \begin{center}
 \includegraphics[height=6.cm]{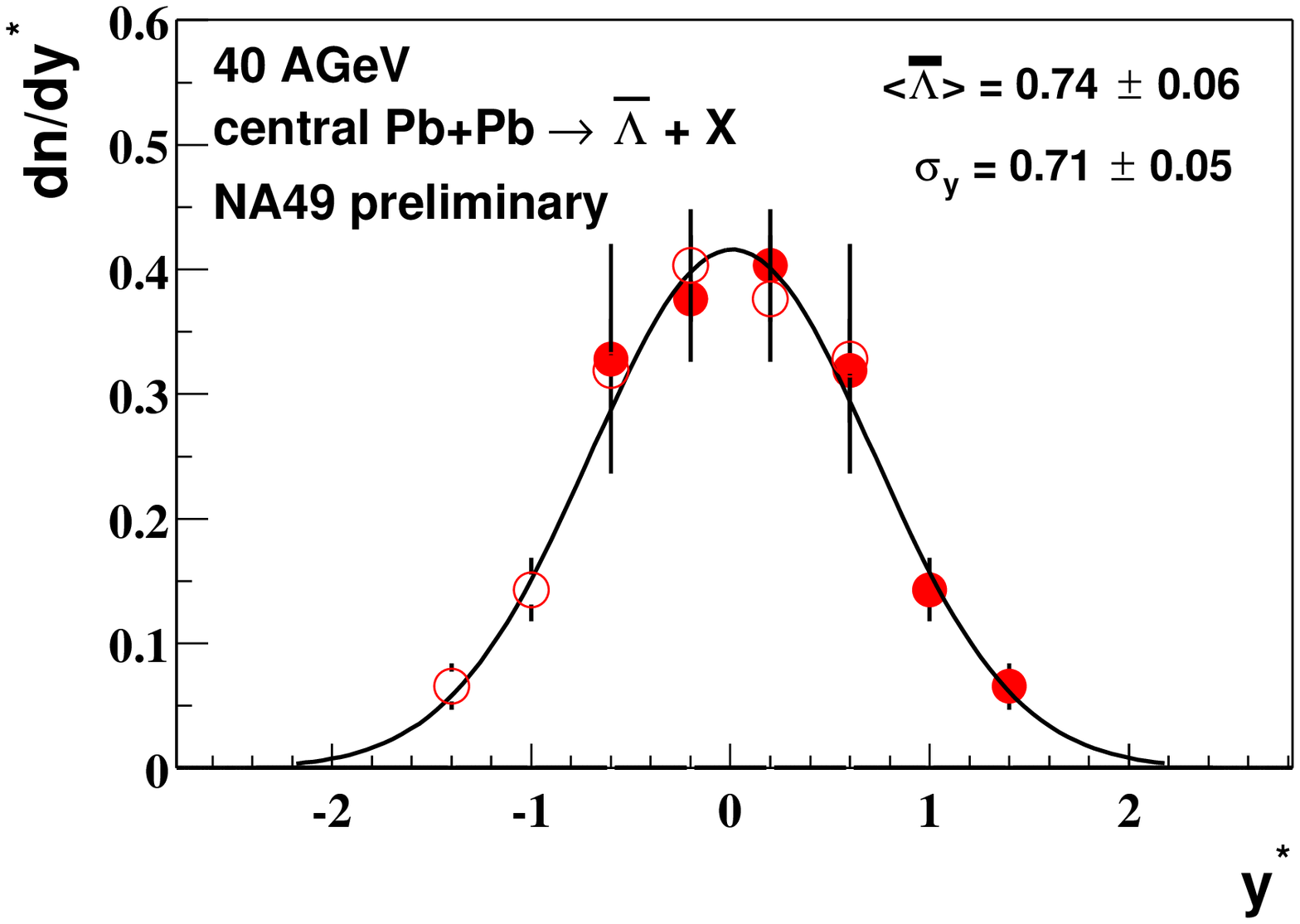}
 \end{center}
\end{minipage} 
\hspace{1cm}
\begin{minipage}[b]{7cm}
 \begin{center}
 \includegraphics[height=6.cm]{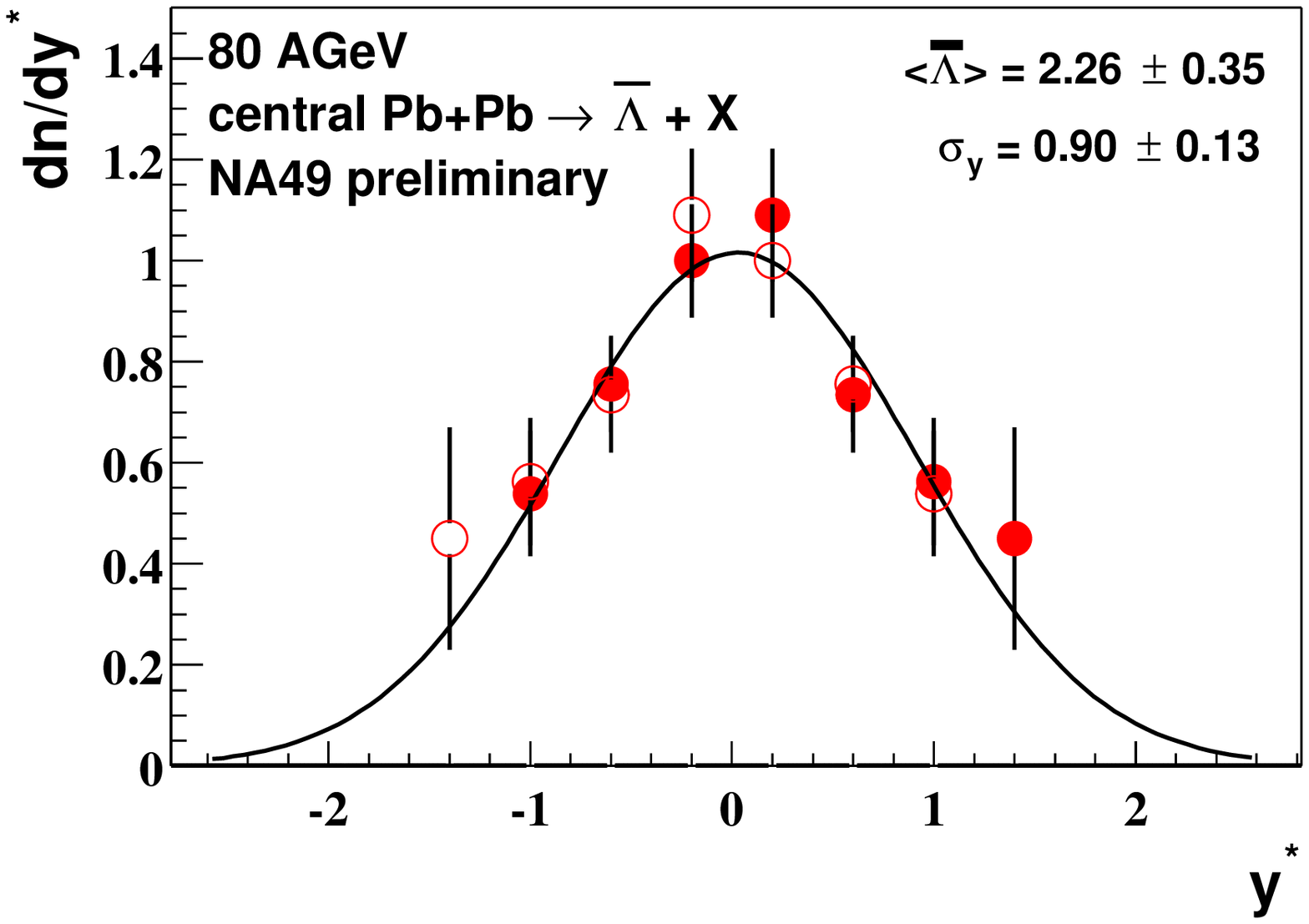}
 \end{center}
\end{minipage} 
 \vspace{-0.8cm}
 \caption{\protect 
The rapidity distribution of $\bar{\Lambda}$ hyperons produced in
central Pb+Pb collisions at 40~(left) and 80~A$\cdot$GeV~(right).} 
 \label{abb8}
\end{figure}

A full analysis, as described in section 3.1., was also performed for
$\bar{\Lambda}$ hyperons. The resulting rapidity spectra are
illustrated in figure~\ref{abb8}. 
The filled circles represent the measured points and the open circles
indicate the points reflected with respect to mid-rapidity.
Both distributions show a maximum at mid-rapidity with a rapidity
density of  
$0.42 \pm 0.03$ and 
$1.06 \pm 0.1$ for 40 and 80~A$\cdot$GeV, respectively. 
The width of a Gaussian fitted to the distribution gives 
$\sigma_{\rm y}=0.71\pm 0.05$ at 40 and 
$\sigma_{\rm y}=0.90 \pm 0.13$ at 80~A$\cdot$GeV. 
The integrated 4$\pi$ yields are 
$0.74 \pm 0.06$ and 
$2.26 \pm 0.35$ $\bar{\Lambda}$ per event for 40 and 80~A$\cdot$GeV,
respectively.  
This means that the rapidity distribution becomes broader going from
40 to 80~A$\cdot$GeV and the maximum increases by a factor of about
2.5.  

The $\bar{\Lambda}/\Lambda$ ratio is studied at mid-rapidity as well
as over the full rapidity range. 
At mid-rapidity this ratio is extracted to be
$\bar{\Lambda}/\Lambda=0.025 \pm 0.0023$ at 40~A$\cdot$GeV, which is
in good agreement with $0.023 \pm 0.001$ 
(uncorrected yields of $\Lambda$ and $\bar{\Lambda}$) measured by the
WA97 collaboration~\cite{WA97_3}, and $\bar{\Lambda}/\Lambda=0.079 \pm
0.01$ at 80~A$\cdot$GeV.  
The $4\pi$ ratios are found to be 
$0.016 \pm 0.0018$ and 
$0.048 \pm 0.005$ for 40 and 80~A$\cdot$GeV, respectively. We deduce
from these numbers that the ratio $\bar{\Lambda}/\Lambda$ increases by
a factor of ~3, at mid-rapidity and $4\pi$, when going from 40 to
80~A$\cdot$GeV. 
That implies a reduction of the baryon chemical potential $\mu_{\rm B}$
of about 27$\%$.  

%
%%%%%%%%%%%%%%%%%%%%%%%%%%%%%%%%%%%%%%%%%%%%%%%%%%%%%%%%%%%%%%%%%%%%%%%%%%%%%%
%
\section{Summary and outlook}

Within the framework of the NA49 energy scan program $\Lambda$ and
$\bar{\Lambda}$ hyperons were measured at 158~A$\cdot$GeV and for the
first time at 40 and 80~A$\cdot$GeV over a large range of rapidity and
transverse momentum. 
The inverse slope parameter near mid-rapidity seems to increase slightly
with increasing energy. This indicates the effect of transverse flow
which is expected to increase with collision energy.   
The rapidity density at mid-rapidity decreases with increasing
energy in the SPS energy range. 
The $\Lambda/\pi$ ratio shows a significant non-monotonic energy
dependence.    
The $\bar{\Lambda}$ rapidity distribution at 80~A$\cdot$GeV is
observed to be broader than at 40~A$\cdot$GeV and the total
multiplicity increases by a factor of ~3. 
The $\bar{\Lambda}/\Lambda$ ratio increases by a factor of ~3, at
mid-rapidity and $4\pi$, when going from 40 to 80~A$\cdot$GeV. 
The planned 20 and 30~A$\cdot$GeV run in the year 2002 will cover the
interesting range between top AGS and low SPS energies.

%%%%%%%%%%%%%%%%%%%%%%%%%%%%%%%%%%%%%%%%%%%%%%%%%%%%%%%%%%%%%%%%%%%%%%%%%%%%%%
%                            Acknowledgements
%%%%%%%%%%%%%%%%%%%%%%%%%%%%%%%%%%%%%%%%%%%%%%%%%%%%%%%%%%%%%%%%%%%%%%%%%%%%%%
\noindent
%{\begin{footnotesize}

\section*{Acknowledgements}
This work was supported by 
the Director, Office of Energy Research, Division of Nuclear Physics
of the Office of High Energy and Nuclear Physics of the US Department
of Energy (DE-ACO3-76SFOOO98 and DE-FG02-91ER40609), 
the US National Science Foundation, 
the Bundesministerium f\"ur Bildung und Forschung, Germany,  
the Alexander von Humboldt Foundation, 
the UK Engineering and Physical Sciences Research Council, 
the Polish State Committee for Scientific Research (5 P03B 13820 and 2
P03B 02418), 
the Hungarian Scientific Research Foundation (T14920 and T23790), 
the EC Marie Curie Foundation, 
and the Polish-German Foundation. 
%\end{footnotesize}}

%%%%%%%%%%%%%%%%%%%%%%%%%%%%%%%%%%%%%%%%%%%%%%%%%%%%%%%%%%%%%%%%%%%%%%%%%%%%%%
%                              References
%%%%%%%%%%%%%%%%%%%%%%%%%%%%%%%%%%%%%%%%%%%%%%%%%%%%%%%%%%%%%%%%%%%%%%%%%%%%%%

%%%%%%%%%%%%%A$\cdot$GeV%%%%%%%%%%%%%%%%%%%%%%%%%%%%%%%%%%%%%%%%%%%%%%%%%%%%%%%%%%%%%%%%%
\end{document}